\documentclass[prd,superscriptaddress,onecolumn,showkeys]{revtex4}
\usepackage{eurosym}
\usepackage{amsfonts}
\usepackage{array}
\usepackage{amsthm}
\usepackage{bm}
\usepackage{palatino}
\usepackage{mathpazo}
\usepackage{amssymb}
\usepackage{eurosym}
\usepackage{amsmath}
\usepackage{epsfig}
\usepackage{graphics}
\usepackage{changes}
\usepackage{color}
\usepackage{graphicx}
\usepackage[colorlinks=true,
            linkcolor=blue,
           urlcolor=black,
           citecolor=black]{hyperref}

\def\be{\begin{equation}}
\def\ee{\end{equation}}
\def\beq{\begin{eqnarray}}
\def\eeq{\end{eqnarray}}

\def\bes{\begin{eqnarray}}
\def\ees{\end{eqnarray}}

\begin{document}
\title{Tunneling Analysis of Kerr-Newman Black Hole-Like Solution in
Rastall Theory}

\author{Riasat Ali}
\email{riasatyasin@gmail.com}
\affiliation{Department of Mathematics, GC University Faisalabad Layyah Campus, Layyah-31200, Pakistan}

\author{Kazuharu Bamba},
\email{bamba@sss.fukushima-u.ac.jp}
\affiliation{Faculty of Symbiotic Systems Science, Fukushima University,
Fukushima 960-1296, Japan}

\author{Syed Asif Ali Shah}
\email{asifalishah695@gmail.com}
\affiliation{Department of Mathematics and Statistics, The University of Lahore 1-Km Raiwind Road,
Sultan Town Lahore 54000, Pakistan}

\author{Muhammad Junaid Saleem}
\email{junaidmathss@gmail.com}
\affiliation{NCBA and E Multan Campus Multan-66000, Pakistan}

\begin{abstract}
Hamilton-Jacobi ansatz is used to analyze boson charged particles tunneling in
Rastall gravity Kerr- Newman black hole (BH) surrounded by perfect fluid matter
through the horizon. The geometrical BH parameters are observed, how these are
affecting the radiation utilizing the Lagrangian gravity equation. In this article,
the corrected Hawking temperature is computed by considering the effects of
quantum gravity. In our analysis, the Rastall gravity BH solution surround
by perfect fluid matter effects on Hawking radiation is analyzed graphically. Moreover,
the stability and instability of Rastall gravity BH are investigated. The
influence of quantum gravity and rotation parameters on BH radiation
are also observed by graphical representation.
\end{abstract}


\date{\today}

\maketitle

\section{Introduction}

Hawking \cite{1} observed that a black hole (BH) radiates all kinds of particles and
the radiated spectrum of particles are absolutely thermal and the spectrum
of emission probability for different types of particles has been studied. The
phenomenon of tunneling is established on the physical action of particles theory which
gives BHs radiation. The law of energy conservation and momentum conservation has been
adapted to this phenomenon. The particular temperature value at which vector particles
are emitted from the BH horizon is called boson radiation. The Hawking radiation for
scalar and fermion particles from different spacetimes have been investigated
\cite{01, 2}. The tunneling probability of charged scalars and fermions outgoing
from the event horizon from these BHs have been studied. The temperature and entropy
in the cosmological horizon of Schwarzschild de-Sitter spacetime have been investigated
in \cite{3}. The charged particles radiated through tunneling from 4-dimensional
as well as 5-dimensional BHs and Vaidya BH have been investigated \cite{5, 6}, however,
the Hawking temperatures and tunneling rates which depend on the properties of
space-time have been observed.

The noncommutative acoustic BHs of quantum-corrected finite entropy have been
discussed \cite{7} and the corrections in terms of electric charge are conserved.
The energy and mass of outgoing fermions and Kerr and Kerr-Newman BHs have been
investigated \cite{8, 9, 10} and the tunneling particles would be stopped at some
particular range and it is also concluded that the tunneling rate of radiated
particles and Hawking temperature for both BHs which is depends on different parameters.
The Hawking radiation spectrum under the influence of quantum gravity has been analyzed
from different types of BH in \cite{x1, x2}.
The energy spectrum and the Hawking temperature contribute holomorphic and
anti-holomorphic functions, GUP and conformal gravity effects on tunneling
radiation from various BHs have been observed \cite{11, 12, 13, 14}, which
are consistent with original results. The BHs tunneling radiation with the
assumption of GUP \cite{15}, unitary theory \cite{16},
loop quantum theory\cite{17} and coordinate system \cite{18} have been examined and found
that which depends on GUP, unitary theory, string and loop quantum theory but independent on
coordinate system.

The tunneling particles from the BH, BTZ like BH and black ring have been examined \cite{19, 20, 21}
and it contains the hypothesis; BHs and black rings have both unstable and stable
properties. Many authors computed
\cite{R1, R2, R3, R4, R5, R6, R7, R8, R9, R10, R11, R11!, R12, R13, R14, R15, R16, R17, R18, R19, R20, R21, R22, R23}
the tunneling radiation of boson and fermion particles to get the Hawking temperature
for different wormholes and BHs. The researcher computed \cite{24, 25} the quantum gravity
effect on tunneling radiation and it is a phenomenon for different types of BHs.

The Newmananis algorithmic rule to analyze the Hayward BH solution with rotation parameter
has been computed in \cite{N1}.The temperature for regular Hayward
BH metric with rotation parameter has been derived with the help of semi-classical phenomenon.
An extension for temperature of 3 and 4 dimensional BH metric as well as higher dimensional BHs metric
under the influences of quantum gravity have been analyzed in \cite{N2, N3}. 
The Rastall rotating BH with surrounding quintessence and dust parameter has been discussed in \cite{N4} and showed that the Rastall theory metric solution is different from the standard metric solution.

The charged bosons particles tunneling with spin equals $1$ that is $Z$ and $W_{\pm}$ play a
significant role for the standard model of electro-weak interaction so that the radiation of
charged boson particles should be importance in the observation of tunneling
radiation. Many different approaches have been analyzed for the investigation of the tunneling spectrum through the
horizons of BHs. Visser \cite{26} analyzed that the Rastall gravity is utterly equivalent to standard Einstein 
general	relativity which is obtained the artificially splitting the physical conserved stress-energy tensor into two non-conserved pieces.

This paper aims to study the quantum gravity effects on
BH thermodynamics. This paper is organized as follows. In the Sec. ({\bf II}) deals the introduction of Rastall gravity Kerr-Newman BH 
surrounded by perfect fluid matter. We analyze the tunneling radiation of BH by
assuming the Lagrangian equation in the scenario of GUP and
calculated the temperature. The graphical representation of temperature in terms
of rotation and quantum gravity are discussed in Sec. ({\bf III}). The last section deals with our results.

\section{Kerr-Newman Black Hole in Rastall Gravity Theory}

By applying Newman-Janis algorithmic rule to study the rotating Rastall BH space-time,
which analyzes by mass ($M$), rotation parameter($a$), surrounding fluid
structure parameter ($\sigma$), state parameter of surrounding fluid $u$ and Rastall coupling parameter $\alpha$.
The Rastall BH metric in the Boyer-Lindquist coordinates from \cite{27, J1} can be expressed as
\begin{equation}
ds^{2}=-Adt^{2}+Bdr^{2}+Cd\theta^{2}
+Dd\phi^{2}+2Fdtd\phi,\label{J}
\end{equation}
where $A$, $B$, $C$,
$D$ and $E$ are given by the following
equations:
\begin{eqnarray}
A&=&1-\frac{2Mr+\sigma r^{\gamma}}{\Sigma},~~~
B=\frac{\Sigma}{\lambda_r},~~~C=\Sigma,\nonumber\\
D&=&sin^{2}\theta
\left(r^{2}+a^{2}+\frac{(2Mr+\sigma r^{\gamma})a^{2}sin^{2}\theta}{\Sigma}\right),\nonumber\\
F&=&\frac{-a sin^{2}\theta\left(2Mr+\sigma r^{\gamma}\right)}{\Sigma}.\nonumber
\end{eqnarray}
Here, these $\lambda_r$, $\Sigma$ and $\gamma$ parameters are defined as
\begin{eqnarray}
\lambda_r&=&r^{2}-2Mr+a^{2}-\sigma r^{\gamma},\nonumber\\
\Sigma &=&a^{2}cos^{2}\theta+a^{2},\nonumber\\
\gamma&=&{\frac{1-3u}{1
-3\alpha(1+u)}}\nonumber
\end{eqnarray}
It is important to mention that, if $a=0$ and $-1<u<\frac{-1}{3}$, the metric (\ref{J}) constitutes
the Kerr BH surrounded by quintessence \cite{J2} and if $\sigma$ and $a$ both are zero, we get
Schwarzschild BH solution in Ref. \cite{J1}.

In order to study the tunneling radiation of charged particles via the BH horizon, we shall analyze electromagnetic
effects of the Lagrangian gravity equation.
The electromagnetic-charged fields, which describe the particle's motion in the Lagrangian charged equation \cite{29}
\begin{equation}
\frac{1}{\sqrt{-g}}\partial_{\mu}(\sqrt{-g}\Psi^{\nu\mu})+\frac{i}{h}e A_{\mu}\Psi^{\nu\mu}+
\frac{m^{2}}{h^{2}}\Psi^{\nu}+
\hbar^{2}\beta\partial_{0}\partial_{0}\partial_{0}(g^{00}\Psi^{0\nu})
-\hbar^{2}\beta\partial_{i}\partial_{i}\partial_{i}(g^{ii}\Psi^{i\nu})=0\label{L},
\end{equation}
where $g$, $\Psi^{\mu\nu}$ and $m$ are the matrix determinant of coefficients,
particle mass and anti-symmetric tensor, respectively.
\begin{equation}
\Psi_{\nu\mu}=(1-\hbar^{2}\beta\partial^{2}_{\nu})\partial{\nu}
\Psi_{\mu}-(1-\hbar^{2}\beta\partial^{2}_{\mu})\partial{\mu}\Psi_{\nu}
+(1-\hbar^{2}\beta\partial^{2}_{\nu})\frac{i}{h}eA_{\nu}\Psi_{\mu}
-(1-\hbar^{2}\beta\partial^{2}_{\mu})\frac{i}{h}eA_{\mu}\Psi_{\nu}.\nonumber
\end{equation}
Here, $\nabla_{\mu}$, $e$ and $A_\mu$  are the covariant derivative, particle charged and BH potential, respectively. So the tunneling of positive and negative
particles is similar but opposite sign $(W_{+}=-W_{-})$. For explanation,
here we shall study the positive particles tunneling and this tunneling
can be changed to negative particles tunneling.
The values of $\Psi^{\mu}$ and
$\Psi^{\nu\mu}$ are given by
\begin{eqnarray}
\Psi^{\nu}&=&\Psi_{\mu}g^{\mu\nu},\nonumber\\
\Psi^{0}&=&\frac{D\Psi_{0}-F\Psi_{3}}{AD+F^{2}},~~~ \Psi^{1}=\frac{\Psi_{1}}{B},~~~
\Psi^{2}=\frac{\Psi_{2}}{C},~~~
\Psi^{3}=\frac{A \Psi_{3}-F \Psi_{0}}{AD+F^{2}},\nonumber\\
\Psi^{pq}&=&\Psi_{\mu\nu} g^{p\mu}g^{q\nu}\nonumber\\
\Psi^{01}&=&\frac{D\Psi_{01}-F\Psi_{31}}{B(AD+F^{2})},~~~
\Psi^{02}=\frac{D\Psi_{02}-F\Psi_{32}}{C(AD+F^{2})},~~~
\Psi^{03}=\frac{-\Psi_{03}}{AD+F^{2}},~~~
\Psi^{12}=\frac{\Psi_{12}}{BC},\nonumber\\
\Psi^{13}&=&\frac{A\Psi_{13}-F\Psi_{10}}{B(AD+F^{2})},~~~\Psi^{23}=\frac{A\Psi_{23}-F\Psi_{20}}{C(AD+F^{2})}.\nonumber
\end{eqnarray}
Applying the WKB approximation,
\begin{equation}
\psi_{\nu}=k_{\nu}\exp[\frac{i}{\hbar}N_{0}(t,r,\theta,\phi)+
\Sigma \hbar^{n}N_{n}(t,r,\theta,\phi)],
\end{equation} The term $\hbar$ is assumed only for the lowest ($1^{st}$) order and
the higher (higher order$>1$) dictate contributions are neglected
in the Lagrangian gravity Eq. (\ref{L}), we get set of field equations.
The variables separation technique in \cite{29}, we can take
\begin{equation}
N_{0}=-E_{0}t+W(r)+V(\theta)+J\phi,
\end{equation}
where $E_{0}=E-\omega\Omega$, $\omega$ and $E$ represent particle's angular momentum and energy
respectively. The set of field equations, we can get a
matrix field equation
\begin{equation*}
K(k_{0},k_{1},k_{2},k_{3})^{T}=0,
\end{equation*}
which implies the "$K$" is label as $4\times4$ matrix, whose elements
are given as follows:
\begin{eqnarray}
K_{00}&=&\frac{D}{B}\dot{W}^{2}W_{1}-\frac{D}{C}J^{2}
J_{1}-\dot{V}^{2}V_{1}-eA_{3}[2\dot{V}+2\beta\dot{V}^{3}+eA_{3}+\beta eA_{3}\dot{V}^{2}]-m^{2}D,\nonumber\\
K_{01}&=&-\frac{D}{B}\dot{W}E_{0}[1+\beta
E_{0}^{2}-eA_{0}E_{0}-
\beta eA_{0}E_{0}]-\frac{F}{B}\dot{W}[\dot{V}+\beta\dot{V}^{3}+eA_{3}+\beta eA_{3}\dot{V}^{2}],\nonumber\\
K_{02}&=&-\frac{D}{C}J[E_{0}-\beta E_{0}^{^{3}}
+eA_{0}+\beta eA_{0}E_{0}]
-\frac{F}{C}J [\dot{V}+\beta\dot{V}^{3}+eA_{3}+\beta\dot{V}^{2}],\nonumber\\
K_{03}&=&\frac{F}{B}\dot{W}^{2}W_{1}+\frac{F}{C}J^{2}J_{1}
-E_{0}E_{1}\dot{V}
+eA_{0}\dot{V}E_{1}-eA_{3}E_{0}
E_{1}+e^{2}A_{0}A_{3}E_{1}+m^{2}F,\nonumber\\
K_{10}&=&-D\dot{W}E_{0}W_{1}
-F\dot{V}\dot{W}W_{1}+DeA_{0}\dot{W}W_{1}-FeA_{3}\dot{W}W_{1},\nonumber\\
K_{11}&=&-DE_{0}^{2}E_{1}
-DeA_{0}E_{0}E_{1}
-F\dot{V}E_{0}V_{1}-FeA_{3}
E_{0}V_{1}
-\frac{F_{0}}{C}J^{2}J_{1}-A\dot{V}^{2}
V_{1}\nonumber\\&&-A_{0}eA_{3}\dot{V}V_{1}
-F\dot{V}E_{0}E_{1}
+FeA_{0}\dot{V}[2+\beta E_{0}^{2}+\beta\dot{V}^{2}]
+DeA_{0}E_{0}E_{1}
-De^{2}A^{2}_{0}E_{1}
\nonumber\\&&+Fe^{2}A_{0}A_{3}V_{1}-AeA_{3}\dot{V}V_{1}
-FeA_{3}E_{0}E_{1}+Fe^{2}A_{0}A_{3}E_{1}-m^{2}F_{0}\nonumber
\end{eqnarray}
\begin{eqnarray}
K_{12}&=&\frac{F_{0}}{C}[\dot{W}J+\beta \dot{W}^{3} J],\nonumber\\
K_{13}&=&F\dot{W}E_{0}
W_{1}+A\dot{W}\dot{V}W_{1}
+AeA_{3}[E_{0}^{2}+\beta\dot{W}^{3}],\nonumber\\
K_{20}&=&-DE_{0}J_{1}-FJ\dot{V}J_{1}
+DeA_{0}JJ_{1}-FeA_{3}JJ_{1},\nonumber\\
K_{21}&=&\frac{F_{0}}{B}\dot{W}JJ_{1},\nonumber\\
K_{22}&=&-DE_{0}^{2}E_{1}
+DeA_{0}E_{0}E_{1}
-F\dot{V}E_{0}V_{1}
-FeA_{3}E_{0}J_{1}
-\frac{F_{0}}{B}[\dot{W}^{2}-\beta \dot{W}^{4}]\nonumber\\&&-\dot{V}^{2}A
V_{1}-AeA_{3}\dot{V}V_{1}
-F\dot{V}E_{0}E_{1}+
FeA_{0}\dot{V}[1+E_{1}+\beta\dot{V}^{2}]\nonumber\\&&+DeA_{0}E_{0}E_{1}-De^{2}A^{2}_{0}E_{1}+Fe^{2}A_{0}A_{3}V_{1}
- AeA_{3}\dot{V}V_{1}
-FeA_{3}E_{0}E_{1}
\nonumber\\&&+Fe^{2}A_{0}A_{3}E_{1}
-Ae^{2}A^{2}_{3}
V_{1}-m^{2}F_{0},\nonumber\\
K_{23}&=&FJE_{0}J_{1}+AJ\dot{V}J_{1}
+AeA_{3}JJ_{1}-FeA_{0}JJ_{1},\nonumber\\
K_{30}&=&-\dot{V}E_{0}V_{1}
-eA_{3}E_{0}V_{1}
+\frac{F}{B}\dot{W}^{2}W_{1}
+\frac{F}{C}J^{2}
J_{1}+eA_{0}\dot{V}V_{1}
+e^{2}A_{0}A_{3}V_{1}+m^{2}F,\nonumber\\
K_{31}&=&\frac{A}{B}\dot{V}\dot{W}V_{1}+
\frac{A}{B}eA_{3}\dot{W}V_{1}+
\frac{F}{B}\dot{W}[E_{0}+\beta E_{0}^{3}-eA_{0}-eA_{0}
\beta E_{0}^{2}],\nonumber\\
K_{32}&=&\frac{A}{C}J\dot{V}V_{1}+
\frac{A}{C}eA_{3}JV_{1}+\frac{F}{C}J[E_{0}
+\beta E_{0}^{3}+eA_{0}+
eA_{0}\beta E_{0}^{2}],\nonumber\\
K_{33}&=&-E_{0}^{2}E_{1}
+eA_{0}E_{0}E_{1}
-\frac{A}{B}\dot{W}^{2}W_{1}-\frac{A}{C}J^{2}J_{1}+eA_{0}
\nonumber\\&&[E_{0}+\beta E_{0}^{3}-eA_{0}-\beta eA_{0}]
-m^{2}A,\nonumber
\end{eqnarray}
where $\dot{W}=\frac{\partial W}{\partial r}$,
$\dot{V}=\frac{\partial V}{\partial\theta}$, $E_{1}=1+\beta E_{0}^{2}$,
$J_{1}=1+\beta J^{2}$, $W_{1}=1+\beta\dot{W}^{2}$,
 $V_{1}=1+\beta\dot{V}^{2}$
and $F_{0}=AD+F^{2}$ and the imaginary parts of $imW_{-}$ and $imW_{+}$ yield
\begin{equation}
imW_{+}=-imW_{-}=\pm \int\sqrt{\frac{(E_{0}-eA_0)^{2}
+X_1[1+\beta\frac{X_2}{X_1}]}{-AB^{-1}}}dr\label{R123}
\end{equation}
Here, $W_{-}$ and $W_{+}$ represent the solution of absorbing and radiating charged boson particles
action respectively and
$X_1=m^{2}A+AC^{-1}J^{2}$ and $X_2=E_{0}^{4}
-eA_{0}E_{0}^{3}+\frac{A}{B}\dot{W}^{4}+\frac{A}{C}J^{4}
-eA_{0}E_{0}^{3}+e^{2}A^{2}_{0}$.
In the series of Taylor's, we are expanding the functions $A(r)$ and $B(r)$ near the horizon, we get
\begin{equation}
A(r_{+})\approx \acute{A}(r)(r-r_{+}),~~B(r_{+})\approx \acute{B}(r)(r-r_{+})\label{w2}
\end{equation}
Since applying Eqs. (\ref{R123}) and (\ref{w2}) and integrate the around the pole, we get
\begin{equation}
imW_{+}=+\pi\iota\frac{(E_{0}-eA_0)^{2}}{2\kappa(r_{+})}(1+\Xi\beta),\label{w1}
\end{equation}
where $\Xi$ is arbitrary parameter and $\kappa(r_+)$ is a
surface gravity of BH. The surface gravity of BH defines as;
\begin{equation}
\kappa(r_{+})=\frac{2M(M-r_{+})}{(a^{2}+a^{2}cos^{2}\theta)^{2}}
\end{equation}
Here, the tunneling probability $\Gamma$ forbidden trajectories of the classically
of the $s$-waves coming from inside to outside from the BH horizon. Applying the
WKB approximation, by the terms of classical action $S_{0}$ of
charged boson particles tunneling across the BH horizon as trajectories up to leading
order is,
\begin{eqnarray}
\Gamma &=&\frac{\Gamma_{emission}}{\Gamma_{absorption}}
=\frac{\exp(-2imW_{+}-2imV)}{\exp(-2imW_{-}-2imV)}
=e^{-4imW_{+}}\nonumber\\&=&\exp\left[-2\pi \frac{(E-\omega\Omega-eA_{0})}{\kappa(r_{+})}(1+\Xi\beta)\right].
\end{eqnarray}
Thus, for computing the Hawking temperature and we expand the
action in terms of particles energy $E$. Hence, the
Hawking temperature is calculated at linear order given by
\begin{equation}
T\cong  \frac{M(r_{+}-M)}{\pi(a^{2}+a^{2}cos^{2}\theta)^{2}}[1-\Xi\beta].
\end{equation}
The charged boson massive particles case is similar as
for massless case for Rastall gravity Kerr-Newman BH
surround by perfect fluid matter from the above Hawking temperature. Moreover,
the temperature for spin-up particles are similar
$(-W_{+}=W_{-})$ as for spin-down case with the
change $r_{+}$ into $r_{-}$. We
observe that both spin-down and spin-up charged boson
particles are radiate with like
rate the temperature is assumed for this case.
\section{Graphical Temperature Analysis}
In this section, we analyze graphical behavior
of thermodynamical quantities in Rastall gravity Kerr-Newman BH surrounded by
perfect fluid matter. For this purpose,
we take a parameter $(\Xi=1)$ and observe the effects
of the thermodynamical quantities (quantum gravity and rotation parameter) on BH radiation. The BH remains unstable
if temperature increases or decreases. From Fig. 1, the Hawking temperature
is decreasing due to the more quantum gravity. The stability of the BH depends on quantum
gravity and also shows that BH remains stable unless
the quantum gravity parameter is zero $(\beta=0)$. Also, temperature
decreases in the approximation range $(0<r_{+}<5)$.\\
If the quantum gravity is assumed to be non-zero, then Hawking
temperature attains the similar value in figure 2. The Hawking
temperature is decrease to rotation parameter in the values 0.1-0.3.
The gravity parameter is fixed and varying the rotation parameter then we concluded
that rotation parameter is inversely proportional to Hawking temperature
and stability of BH. The stability of BH remains constant in very small
range, i.e., approximation $0<r_{+}<0.5$ and then
temperature decreases with a wide variation.
The rotation parameter is inversely proportional to the temperature of BH by keeping
the quantum gravity constant in Fig. 2 for 2D.
The temperature is behaving constant when
$r_{+}$ is very small, but at a certain value of
rotation parameter, it attain a constantly decrease value in Fig. 2 for 2D.
Hence our BH is stable when $r_{+}$ is very small range otherwise is unstable.

We concluded that BH radiation depends on BH~~mass, rotation parameter and gravity parameter
but keeping constant $cos^{2}\theta=1$.\\

\begin{figure}\begin{center}
\epsfig{file=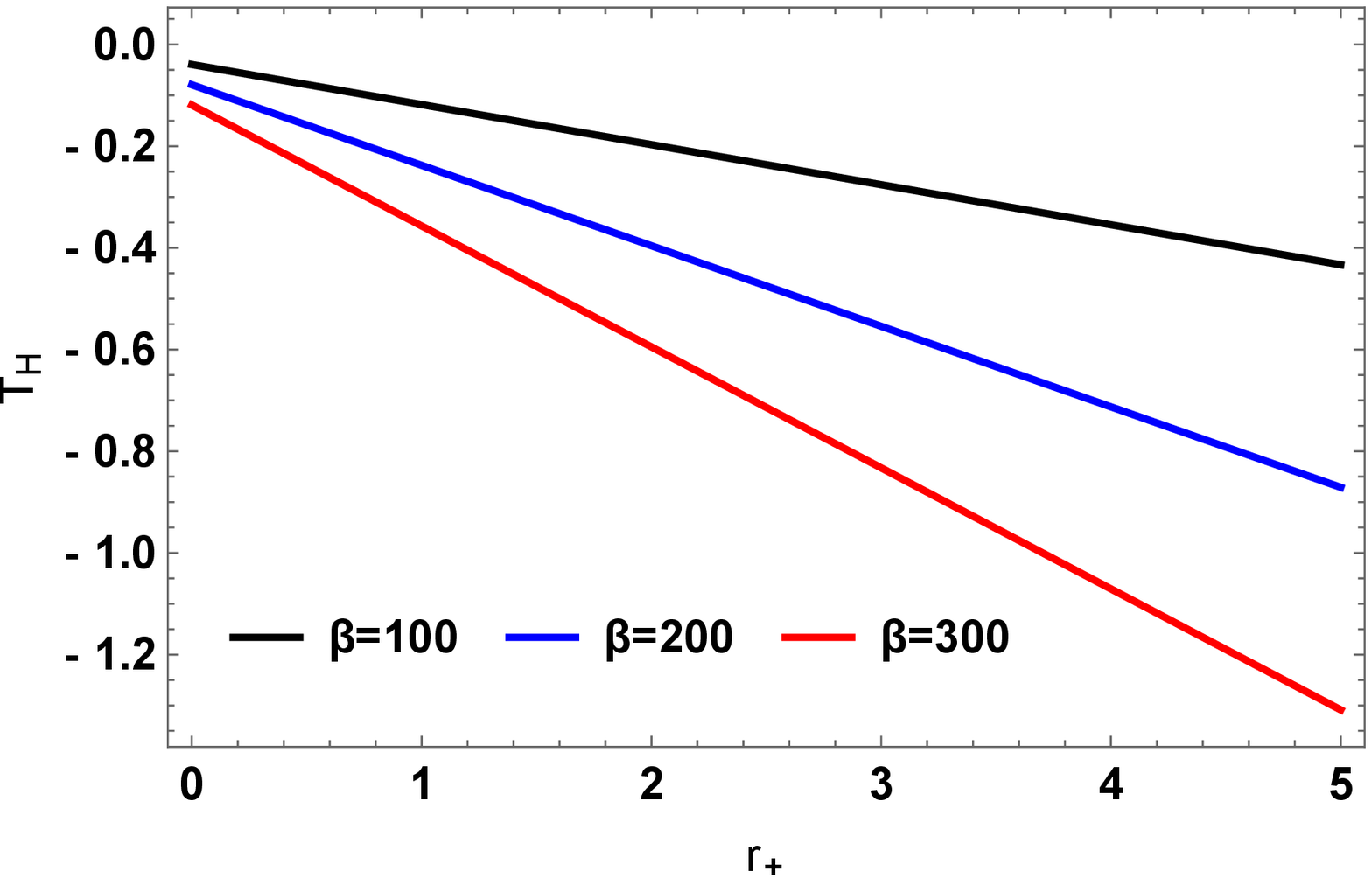, width=0.60\linewidth}
\caption{$T_{H}$ versus $r_{+}$ for $M>0,~a=10,~~cos^{2}\theta=1$ and $\Xi=1.$}
\end{center}
\begin{center}
\epsfig{file=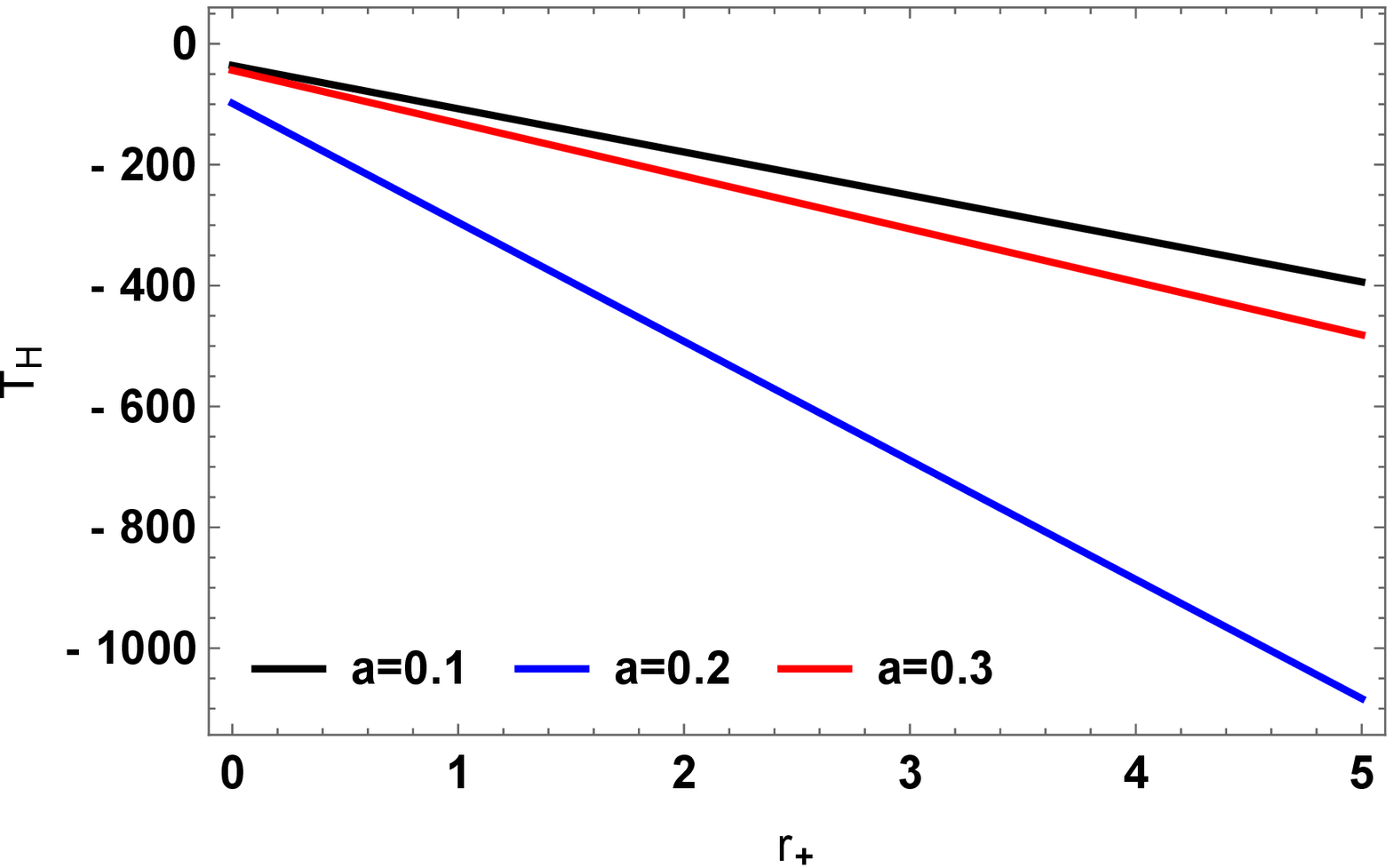, width=0.60\linewidth}
\caption{$T_{H}$ versus $r_{+}$ for $M>0,~~\beta>0,~~cos^{2}\theta=1$ and $\Xi=1.$}
\end{center}
\end{figure}

\section{Conclusions}

In this paper, we have observed radiation spectrum by considering Hawking
temperature for Rastall gravity Kerr-Newman BH is surrounded by perfect
fluid matter. For this purpose, we have utilized the WKB approximation to
the Hamilton-Jacobi ansatz for massive bosons charged spin-1 particles.
The calculation yields, the Hawking temperature reliable with BH universality.
In our investigation, we have changed the Lagrangian field equation in curved space time
by effect of incorporate GUP quantum gravity. We have computed
the tunneling probability at event horizon as well as the equating Hawking
temperature. We have observed temperature in detail the effect of quantum gravity and rotation parameters in graphically.

The paper have studied graphical investigation of the
result and the BH will emit away all types of particles as the black-body
radiation and the BH lose completely its information. In adjustment to break this
problem, the after-effect rotation and gravity to be modified. We assume into account the law of
conservation of momentum and energy. In our observation, the back reaction
of the radiated boson particles of the self-gravitational interaction and BH geometry are
reasonably ignored, the calculated Hawking temperatures depend
on quantum gravity and the leading terms of metric.
The modified form of temperature depends on BH~~mass,
rotation parameters and quantum gravity parameter.
When the gravity parameter $(\beta=0)$ influences are ignored, we have
obtained the temperature of Rastall gravity Kerr-Newman BH.
So, the quantum gravity effect minimum then stability of BH are
increased.

The quantum gravity and rotation are present in the BH,
which shows that the instability of BH.
The BH is completely stable when rotation and quantum gravity do not appear.\\
{\bf Acknowledgments}\\
The work of KB was supported in part by the JSPS KAKENHI Grant Number JP21K03547.

\section*{Appendix}

In the WKB approximation the term $\hbar$ is considered just for the $1^{st}$ order and
the higher dictate contributions are neglected
in the Lagrangian Eq. (\ref{L}), we get following field equations;
\begin{eqnarray}
&&\frac{D}{B}[k_{1}(\partial_{0}N_{0})(\partial_{1}N_{0})
+\beta k_{1}(\partial_{0}N_{0})^{3}(\partial_{1}N_{0})
+k_{0}(\partial_{1}N_{0})^{2}-\beta k_{0}(\partial_{1}N_{0})^{4}+eA_{0}k_{1}(\partial_{0}N_{0})^{2}(\partial_{1}N_{0})+
\beta eA_{0}k_{1}(\partial_{0}N_{0})^{2}(\partial_{1}N_{0})]\nonumber\\
&&-\frac{F}{B}
[k_{1}(\partial_{1}N_{0})(\partial_{3}N_{0})
+\beta k_{1}(\partial_{1}N_{0})(\partial_{3}N_{0})^{3}
-k_{3}(\partial_{1}N_{0})^{2}-\beta k_{3}(\partial_{1}N_{0})^{4}
+eA_{3}k_{1}(\partial_{1}N_{0})
+\beta eA_{3}k_{1}(\partial_{1}N_{0})(\partial_{3}N_{0})^{2}]\nonumber\\
&&+\frac{D}{C}[k_{2}(\partial_{0}N_{0})(\partial_{2}N_{0})
+\beta k_{2}(\partial_{0}N_{0})^{3}(\partial_{2}N_{0})
-k_{o}(\partial_{2}N_{0})^{2}-\beta k_{0}(\partial_{2}N_{0})^{4}
+eA_{0}k_{2}(\partial_{2}N_{0})+\beta eA_{0}
k_{2}(\partial_{0}N_{0})^{2}(\partial_{2}N_{0})]\nonumber\\
&&-\frac{F}{C}[k_{2}(\partial_{2}N_{0})(\partial_{3}N_{0})+\beta k_{2}
(\partial_{2}N_{0})(\partial_{3}N_{0})^{3}-k_{3}
(\partial_{2}N_{0})^{2}-\beta k_{3}(\partial_{2}N_{0})^{4}
+eA_{3}k_{2}(\partial_{2}N_{0})+\beta eA_{3}k_{2}
(\partial_{2}N_{0})(\partial_{3}N_{0})^{2}]
\nonumber\\
&&+[k_{3}(\partial_{0}N_{0})+\beta k_{3}(\partial_{0}N_{0})^{3}
(\partial_{3}N_{0})
-k_{0}(\partial_{3}N_{0})^{2}-\beta k_{0}
(\partial_{3}N_{0})^{4}+eA_{0}k_{3}(\partial_{3}N_{0})
+\beta eA_{0}k_{3}(\partial_{0}N_{0})^{2}(\partial_{3}N_{0})-eA_{3}k_{0}(\partial_{3}N_{0})\nonumber\\
&&-\beta eA_{3}k_{0}
(\partial_{3}N_{0})^{3}]+eA_{3}[k_{3}
(\partial_{0}N_{0})+\beta k_{3}(\partial_{0}N_{0})^{3}
-k_{0}(\partial_{3}N_{0})-\beta k_{0}(\partial_{3}N_{0})^{3}
+eA_{0}k_{3}+\beta eA_{0}k_{3}(\partial_{0}N_{0})^{2}
-eA_{3}k_{0}\nonumber\\
&&-\beta eA_{3}k_{0}(\partial_{3}N_{0})^{2}]=0,\\
&&D[k_{0}(\partial_{0}N_{0})(\partial_{1}N_{0})+
\beta k_{0}(\partial_{0}N_{0})(\partial_{1}N_{0})^{3}
-k_{1}(\partial_{0}N_{0})^{2}-\beta k_{1}(\partial_{0}N_{0})^{4}-eA_{0}k_{1}(\partial_{0}N_{0})-\beta eA_{0}k_{1}(\partial_{0}N_{0})^{3}]-F
[k_{3}(\partial_{0}N_{0})\nonumber\\
&&(\partial_{1}N_{0})+\beta k_{3}(\partial_{0}N_{0})(\partial_{1}N_{0})^{3}
-k_{1}(\partial_{0}N_{0})(\partial_{3}N_{0})-\beta k_{1}
(\partial_{0}N_{0})(\partial_{3}N_{0})^{3}
-eA_{3}k_{1}(\partial_{0}N_{0})
-\beta eA_{3}k_{1}(\partial_{0}N_{0})(\partial_{3}N_{0})^{2}]
\nonumber\\
&&+(AD-F^{2})C^{-1}[k_{2}(\partial_{1}N_{0})(\partial_{2}N_{0})
+\beta k_{2}(\partial_{1}N_{0})^{3}(\partial_{2}N_{0})
-k_{1}(\partial_{2}N_{0})^{2}-\beta k_{1}(\partial_{2}N_{0})^{4}]
+A[k_{3}(\partial_{1}N_{0})(\partial_{3}N_{0})+\beta k_{3}\nonumber\\
&&(\partial_{1}N_{0})^{3}
(\partial_{3}N_{0})
-k_{1}(\partial_{3}N_{0})^{2}-\beta k_{1}(\partial_{3}N_{0})^{4}
-eA_{3}k_{1}(\partial_{3}N_{0})-\beta eA_{3}k_{1}(\partial_{3}N_{0})^{3}]
-F[k_{0}(\partial_{1}N_{0})(\partial_{3}N_{0})+\beta k_{0}
(\partial_{1}N_{0})^{3}\nonumber\\
&&(\partial_{3}N_{0})-k_{1}(\partial_{0}N_{0})
(\partial_{3}N_{0})-\beta k_{1}(\partial_{0}N_{0})^{3}(\partial_{3}N_{0})
-eA_{0}k_{1}(\partial_{3}N_{0})-\beta eA_{0}k_{1}(\partial_{0}N_{0})^{2}(\partial_{3}N_{0})]
+D eA_{0}[k_{0}(\partial_{1}N_{0})+\beta k_{0}\nonumber\\
&&(\partial_{1}N_{0})^{3}-k_{1}(\partial_{0}N_{0})-\beta k_{1}(\partial_{0}N_{0})^{3}
-eA_{0}k_{1}
-\beta eA_{0}k_{1}(\partial_{0}N_{0})^{2}]-F eA_{0}[k_{3}(\partial_{1}N_{0})+\beta k_{3}(\partial_{1}N_{0})^{3}-k_{1}(\partial_{3}N_{0})
-\beta k_{1}\nonumber\\
&&(\partial_{3}N_{0})^{3}-eA_{3}k_{1}
-\beta eA_{3}k_{1}(\partial_{3}N_{0})^{2}]+eAA_{3}[k_{3}(\partial_{0}N_{0})^{2}+\beta k_{3}
(\partial_{1}N_{0})^{3}-k_{1}(\partial_{3}N_{0})
-\beta k_{1}(\partial_{3}N_{0})^{3}-eA_{3}k_{1}-\beta eA_{3}\nonumber\\
&&k_{1}(\partial_{3}N_{0})^{2}]-
eFA_{3}[k_{0}(\partial_{1}N_{0})+\beta k_{0}(\partial_{1}N_{0})^{3}-
k_{1}(\partial_{0}N_{0})
-\beta k_{1}(\partial_{0}N_{0})^{3}-eA_{0}k_{1}-\beta eA_{0}k_{1}
(\partial_{0}N_{0})^{2}]-m^{2}k_{1}\nonumber\\
&&(AD-F^{2})=0\\
&&D[k_{0}(\partial_{0}N_{0})(\partial_{2}N_{0})+\beta k_{0}
(\partial_{0}N_{0})(\partial_{2}N_{0})^{3}
-k_{2}(\partial_{0}N_{0})^{2}-\beta k_{2}(\partial_{0}N_{0})^{4}
-eA_{0}k_{2}(\partial_{0}N_{0})-\beta eA_{0}k_{2}(\partial_{0}N_{0})^{3}]-F
[k_{3}(\partial_{0}N_{0})\nonumber\\
&&(\partial_{2}N_{0})+\beta k_{3}(\partial_{0}N_{0})(\partial_{2}N_{0})^{3}-k_{2}(\partial_{0}N_{0})
(\partial_{3}N_{0})-\beta k_{2}(\partial_{0}N_{0})(\partial_{3}N_{0})^{3}
-eA_{3}k_{2}(\partial_{0}N_{0})-\beta eA_{3}k_{2}(\partial_{0}N_{0})(\partial_{3}N_{0})^{2}]\nonumber\\
&&+(AD-F^{2})B^{-1}[k_{1}(\partial_{1}N_{0})(\partial_{2}N_{0})
+\beta k_{1}(\partial_{1}N_{0})(\partial_{2}N_{0})^{3}
-k_{2}(\partial_{1}N_{0})^{2}-\beta k_{2}(\partial_{1}N_{0})^{4}]
+A[k_{3}(\partial_{2}N_{0})(\partial_{3}N_{0})+\beta k_{3}\nonumber\\
&&(\partial_{2}N_{0})^{3}
(\partial_{3}N_{0})
-k_{2}(\partial_{3}N_{0})^{2}-\beta k_{3}(\partial_{3}N_{0})^{4}-eA_{3}k_{2}
(\partial_{3}N_{0})-\beta eA_{3}k_{2}(\partial_{3}N_{0})^{3}]
-F[k_{0}(\partial_{2}N_{0})(\partial_{3}N_{0})+\beta k_{0}
(\partial_{2}N_{0})^{3}\nonumber\\
&&(\partial_{3}N_{0})-eA_{0}k_{2}(\partial_{3}N_{0})
-\beta eA_{0}k_{2}(\partial_{0}N_{0})^{2}(\partial_{3}N_{0})
-k_{2}(\partial_{0}N_{0})(\partial_{3}N_{0})-\beta k_{2}
(\partial_{0}N_{0})^{3}(\partial_{3}N_{0})]
+DeA_{0}[k_{0}(\partial_{2}N_{0})+\beta k_{0}\nonumber\\
&&(\partial_{2}N_{0})^{3}-k_{2}(\partial_{0}N_{0})-\beta k_{2}(\partial_{0}N_{0})^{3}-eA_{0}k_{2}
-\beta eA_{0}k_{2}(\partial_{0}N_{0})^{2}]-FeA_{0}[k_{3}(\partial_{2}N_{0})
+\beta k_{3}(\partial_{2}N_{0})^{3}-k_{2}(\partial_{3}N_{0})
-\beta k_{2}\nonumber\\
&&(\partial_{3}N_{0})^{3}-eA_{3}k_{2}
-\beta eA_{3}k_{2}(\partial_{3}N_{0})^{2}]
+eAA_{3}[k_{3}(\partial_{2}N_{0})+\beta k_{3}
(\partial_{2}N_{0})^{3}-k_{2}(\partial_{3}N_{0})
-\beta k_{2}(\partial_{3}N_{0})^{3}\nonumber\\
&&-eA_{3}k_{2}-\beta eA_{3}k_{2}(\partial_{3}N_{0})^{2}]-eFA_{3}
[k_{0}(\partial_{2}N_{0})+\beta k_{0}(\partial_{2}N_{0})^{3}-
k_{2}(\partial_{0}N_{0})
-\beta k_{2}(\partial_{0}N_{0})^{3}-eA_{0}k_{2}-\beta eA_{0}k_{2}
(\partial_{0}N_{0})^{2}]\nonumber\\
&&-m^{2}k_{2}(AD-F^{2})=0\\
&&[k_{0}(\partial_{0}N_{0})(\partial_{3}N_{0})+\beta k_{0}
(\partial_{0}N_{0})(\partial_{3}N_{0})^{3}
+k_{3}(\partial_{0}N_{0})^{2}-\beta k_{3}(\partial_{0}N_{0})^{4}
+eA_{3}k_{0}(\partial_{0}N_{0})+\beta eA_{3}k_{0}
(\partial_{0}N_{0})(\partial_{3}N_{0})^{2}\nonumber\\
&&-eA_{0}k_{3}(\partial_{0}N_{0})-\beta eA_{0}k_{3}
(\partial_{0}N_{0})^{3}]
+\frac{A}{B}
[k_{1}(\partial_{1}N_{0})(\partial_{3}N_{0})+\beta k_{1}
(\partial_{1}N_{0})(\partial_{3}N_{0})^{3}
-k_{3}(\partial_{1}N_{0})^{2}-\beta k_{3}(\partial_{1}N_{0})^{4}
+\nonumber\\
&&eA_{3}k_{1}(\partial_{1}N_{0})+\beta eA_{3}k_{1}(\partial_{1}N_{0})(\partial_{3}N_{0})^{2}]
-\frac{F}{B}[k_{1}(\partial_{0}N_{0})(\partial_{1}N_{0})
-\beta k_{1}(\partial_{0}N_{0})^{3}(\partial_{1}N_{0})
-k_{o}(\partial_{1}N_{0})^{2}-\beta k_{0}(\partial_{1}N_{0})^{4}
\nonumber\\
&&+eA_{0}k_{1}(\partial_{1}N_{0})+\beta eA_{0}
k_{1}(\partial_{0}N_{0})^{2}(\partial_{1}N_{0})]
+\frac{A}{C}[k_{2}(\partial_{2}N_{0})(\partial_{3}N_{0})+\beta k_{2}
(\partial_{2}N_{0})(\partial_{3}N_{0})^{3}-k_{3}(\partial_{2}N_{0})^{2}-
\beta k_{3}(\partial_{2}N_{0})^{4}\nonumber\\
&&+eA_{3}k_{2}(\partial_{2}N_{0})+\beta eA_{3}k_{2}
(\partial_{2}N_{0})(\partial_{3}N_{0})^{2}]
-\frac{F}{C}[k_{2}(\partial_{0}N_{0})(\partial_{2}N_{0})+
\beta k_{2}(\partial_{0}N_{0})^{2}(\partial_{2}N_{0})
-k_{0}(\partial_{2}N_{0})^{2}-\beta k_{0}(\partial_{2}N_{0})^{4}
\nonumber\\
&&+eA_{0}k_{2}(\partial_{2}N_{0})
+\beta eA_{0}k_{2}(\partial_{0}N_{0})^{2}(\partial_{2}N_{0})]
+eA_{0}[k_{0}
(\partial_{3}N_{0})+\beta k_{0}(\partial_{3}N_{0})^{3}-k_{3}(\partial_{0}N_{0})-\beta k_{3}(\partial_{0}N_{0})^{3}
+eA_{3}k_{0}+\nonumber\\
&&\beta eA_{3}k_{0}(\partial_{3}N_{0})^{2}
-eA_{0}k_{3}-\beta eA_{0}k_{3}(\partial_{0}N_{0})^{2}]+m^{2}
[Fk_{0}-Ak_{3}]=0.
\end{eqnarray}

\end{document}